\documentclass[doublecol]{epl2} 

\usepackage{psfrag,graphicx}
\usepackage{dcolumn}
\usepackage{bm}
\usepackage{amsfonts,amssymb,amsmath}        
\usepackage{amssymb}
\usepackage{xcolor}

\title{On the Emergent Dynamics of Fermions in Curved Spacetime}

\author{Giandomenico Palumbo\inst{1}}
\shortauthor{Giandomenico Palumbo}

\institute{                    
  \inst{1} Institute for Theoretical Physics, Center for Extreme Matter and Emergent Phenomena,
Utrecht University,\\ Leuvenlaan 4, 3584 CE Utrecht, The Netherlands
}

\abstract{
Relativistic spin-1/2 particles in curved spacetime are naturally described by Dirac theory, which is a dynamical and Lorentz-invariant field theory. In this work, we propose a non-dynamical fermion theory in 3+1 dimensions dubbed spinor topological field theory, built in terms of a spinor field and a Cartan connection related to de Sitter group. We show that our model gives rise to the Dirac theory in curved spacetime when the local de Sitter gauge invariance of the model breaks down to the Lorentz gauge invariance, providing also a geometric origin to the fermion mass.
Finally, we show that other gauge fields and suitable four-fermion interactions can be included in a straightforward way.}

\begin{document}

\maketitle

\section{Introduction}
\label{intro}
The Dirac theory of fermions plays a central role in modern physics. Relativistic fermions are not only the main constituents of matter at fundamental level but they also emerge as quasiparticles in several condensed matter systems like graphene, topological insulators and superconductors \cite{Ryu}. The generalization of Dirac theory in curved spacetime was proposed by Fock and Ivanenko in Ref. \cite{Fock}. They introduced the concept of spin connection and tetrads which allow us to define the Dirac operator on a generic spin-manifold. Tetrads are directly related to the metric tensor, while the spin connection comes from the local covariance of spinor field with respect to the Lorentz group. When spinors are integrated out in the corresponding partition functions, then suitable topological terms can appear in the corresponding effective actions \cite{Palumbo-Index}. Around the ground state, these terms are dominant with respect to the non-topological ones. This implies that topological field theories (TFTs) describe the physical properties of Dirac particles coupled to gauge fields and curved background in the low-energy regime. By definition, TFTs are non-dynamical gauge theories, i.e. there are no local propagating degrees of freedom on shell because their actions are metric-independent \cite{Birmingham}. However, they have important applications in topological phases of matter \cite{Pachos}, high-energy physics and gravity \cite{Baez,Freidel}. \\
The goal of this paper is to provide and analyze a new kind of femion theory in 3+1 dimensions dubbed spinor topological field theory, built in terms of a spinor field and a Cartan connection related to the de Sitter group \cite{Wise,Randono}. We show that this gauge theory gives rise to the standard Dirac theory in curved spacetime when the de Sitter gauge invariance breaks down to the Lorentz gauge invariance. This means that the dynamics of fermions emerges in the low energy limit, after a phase transition induced by a (spontaneous) symmetry breaking.
Within this framework, the model provides also a geometric origin to the fermion mass, showing also some connections with the gravitational Einstein-Cartan theory \cite{Hehl}, formulated through the Cartan-geometry language. Finally, we show that other gauge fields and suitable four-fermion interactions can be naturally included.

\section{Spinor topological field theory}
\label{Spinor topological field theory}
We start considering a gauge theory in a 3+1-dimensional Lorentzian spacetime $M_{4}$, defined by the following action 
\begin{eqnarray}\label{spinor-BF}
S_{\psi TFT}\left[\overline{\Psi},\Psi, A_{\mu}\right]=\hspace{2.0cm}   \\ \nonumber=\int_{M_{4}} d^{4}x\,
\,\epsilon^{\mu\nu\alpha\beta}\,\left(\vartheta\,\overline{D_{\mu}^{A}\Psi}\,F_{\nu\alpha}\,D_{\beta}^{A}\Psi+\theta\,\overline{\Psi}F_{\mu\nu}F_{\alpha\beta}\Psi\right), 
\end{eqnarray}
where $\epsilon^{\mu\nu\alpha\beta}$ is the Levi-Civita symbol, $\mu=0,1,2,3$, $\vartheta=i/3$, $\theta=-\frac{i}{4!}\zeta$, with $\zeta$ a dimensionless parameter, and $\Psi$ is a (dimensionless) four-component spinor. Here, $D_{\mu}^{A}=\partial_{\mu}+A_{\mu}$ and $F_{\mu\nu}=\left[D_{\mu}^{A}, D_{\nu}^{A}\right]$ are the covariant derivative and curvature tensor, respectively, of a connection $A_{\mu}$ that takes values in the $spin(4,1)$ algebra, i.e. the double covering of the $so(4,1)$ de Sitter algebra \cite{Randono} (clearly, $\overline{D_{\mu}^{A}\Psi}=\partial_{\mu}\overline{\Psi}+\overline{\Psi} A_{\mu}$).
This theory is gauge invariant because the curvature tensor and the covariant derivative transform as follows
\begin{eqnarray}\label{gaugeinv}
F_{\mu\nu}\rightarrow U(x)F_{\mu\nu}U(x)^{\dagger}, \hspace{0.3cm} D^{A}_{\alpha}\rightarrow U(x) D^{A}_{\alpha}U(x)^{\dagger},
\end{eqnarray}
while the spinor field $\Psi$ transforms covariantly with respect to the $Spin(4,1)$ group, i.e.
\begin{eqnarray}\label{gaugeinv2}
\Psi\rightarrow U(x) \Psi, \hspace{0.3cm} \overline{\Psi}\rightarrow \overline{\Psi} U(x)^{\dagger},
\end{eqnarray} 
where $U(x)$ in both Eqns. (\ref{gaugeinv})  and (\ref{gaugeinv2}) is a local matrix belonging to $Spin(4,1)$ and $\dagger$ corresponds to the Krein adjoint (see, for instance Ref. \cite{Walter}).
The representation of this group is given in terms of suitable products of $4 \times 4$ Dirac matrices $\{\gamma_{a}, \gamma_{5}\}$, \cite{Randono2}, with $a=0,1,2,3$, while $\gamma_{5}=i\gamma_{0}\gamma_{1}\gamma_{2}\gamma_{3}$ is the chiral matrix.\\
Importantly, the action (\ref{spinor-BF}) is metric independent and describes a topological field theory. In other words, the fermion field $\Psi$ has no local propagating degrees of freedom because the Dirac operator $\displaystyle{\not}D=i\,e^{\mu}_{a}\gamma^{a}D_{\mu}$, where $e^{\mu}_{a}$ are the co-tetrads, is absent in Eqn. (\ref{spinor-BF}). In fact, the metric tensor can be defined in terms of the tetrads $e_{\mu}^{a}$, namely 
\begin{eqnarray}
g_{\mu\nu}=e_{\mu}^{a}\,e_{\nu}^{b}\,\eta_{ab},
\end{eqnarray} 
 where the local flat metric $\eta_{ab}$ can be written trough the anti-commutation relations between the Dirac matrices, i.e. $\{\gamma_{a},\gamma_{b}\}=2\,\eta_{ab}\mathbb{I}$, where $\mathbb{I}$ is the identity matrix. In alternative, a dynamical theory can be characterized by an action that contains an internal Hodge dual operator \cite{Wise}, but this is not the case for the action (\ref{spinor-BF}).\\
At this point, one can wonder if there is any relation between our model and the Dirac theory, although the latter describes dynamical fermions. Moreover, the de Sitter group seems to play no fundamental role in the Standard Model. In fact, in the curved spacetime, the fermion field transforms covariantly under the local $Spin(3,1)$ group which is the double covering of the Lorentz group $SO(3,1)$. In this case, the corresponding covariant derivative $D^{\omega}_{\mu}=\partial_{\mu}+\omega_{\mu}$ is built through the spin connection $\omega_{\mu}=(i/2)\omega_{\mu}^{ab}\Sigma_{ab}$, where $\Sigma_{ab}=\,i\,[\gamma_{a},\gamma_{b}]/4$ are the generators of $Spin(3,1)$.\\
We now show that the Dirac action can be naturally derived from our action (\ref{spinor-BF}) through a (spontaneous) symmetry breaking of the de Sitter gauge invariance, such that only the gauge invariance of its Lorentz subgroup survives in the theory. In other words, the original gauge transformations of $\Psi$ in (\ref{gaugeinv2}), reduce to
\begin{eqnarray}\label{gaugeinv3}
\Psi\rightarrow \widehat{U}(x) \Psi, \hspace{0.3cm} \overline{\Psi}\rightarrow \overline{\Psi} \widehat{U}(x)^{\dagger},
\end{eqnarray}
where now $\widehat{U}(x)$ belongs to $Spin(3,1)$. \\In this work, we do not describe the exact nature of the underlying Higgs-like mechanism but we note that suitable symmetry-breaking mechanisms for the de Sitter group have been already proposed in pure gravity \cite{Stelle, Randono2,Westman} and in even in presence of fermions \cite{Ikeda}.\\
In this context, $A_{\mu}$ has to be considered a Cartan connection \cite{Wise,Randono}. This connection plays a central role in Cartan geometry which represents a generalization of Riemannian geometry, where the flat tangent space is replaced by a curved one \cite{Randono}. The Cartan connection is then related to a generalized tangent space equivalent, in our case, to the de Sitter space. Among several applications, this kind of differential geometry has been successfully employed in the study of 2+1-dimensional topological gravity \cite{Wise,Wise2}. Importantly, in Cartan geometry, there exists a unique decomposition (up to a sign, \cite{Wise2}) of $A_{\mu}$ that is given by
\begin{eqnarray}\label{decomposition}
A_{\mu}=\omega_{\mu}-\frac{i}{2\,l}\,\gamma_{5}\gamma_{a}\,e_{\mu}^{a},
\end{eqnarray}
where $l$ is a dimensionful constant parameter related to the inverse of the radius of the de Sitter tangent space. In this way, the corresponding curvature tensor $F_{\mu\nu}$ is decomposed as follows
\begin{eqnarray}\label{decomposition2}
F_{\mu\nu}=R_{\mu\nu}^{\omega}-\frac{1}{4\,l^{2}}(e_{\mu}^{a}e_{\nu}^{b}-e_{\nu}^{a}e_{\mu}^{b})[\gamma_{a},\gamma_{b}]-\frac{i}{2\,l}\gamma_{5}\gamma_{a}\,T^{a}_{\mu\nu},
\end{eqnarray}
where $R_{\mu\nu}^{\omega}=R_{\mu\nu}^{ab}\,i\,[\gamma_{a},\gamma_{b}]/4$ and $T_{\mu\nu}^{a}$ are the Riemann and torsion tensors, respectively.
We have now all the ingredients necessary to derive the Dirac theory. We have just to rewrite $A_{\mu}$ and $F_{\mu\nu}$ in (\ref{spinor-BF}) in terms of $\omega_{\mu}$ and $e_{\mu}^a$ through Eqns. (\ref{decomposition}) and (\ref{decomposition2}). 
In order to simplify the calculations, let us define the following variables $\widehat{\gamma}_{\mu}=e_{\mu}^{a}\gamma_{a}$, being the metric tensor $g_{\mu\nu}$ on $M_4$ identified through their anti-commutation relations, i.e.
\begin{eqnarray}
\{\widehat{\gamma}_{\mu},\widehat{\gamma}_{\nu}\}=2 g_{\mu\nu}\mathbb{I}.
\end{eqnarray}
Here, we focus on the terms in the action (\ref{spinor-BF}) which are multiplied only by tetrads. The corresponding Lagrangian density for these terms, is given by
 \begin{align}\label{equations}
\epsilon^{\mu\nu\alpha\beta}\overline{\Psi}\left(-\frac{1}{12\,l^{3}}\gamma_{5}\widehat{\gamma}_{\mu}[\widehat{\gamma}_{\nu},\widehat{\gamma}_{\alpha}]D_{\beta}^{A}+\frac{\theta}{4\,l^{4}}[\widehat{\gamma}_{\mu},\widehat{\gamma}_{\nu}][\widehat{\gamma}_{\alpha},\widehat{\gamma}_{\beta}]\right)\Psi=\nonumber \\ -\frac{\epsilon^{\mu\nu\alpha\beta}}{3!\,l^{3}}\,\overline{\Psi}\,\gamma_{5}
\widehat{\gamma}_{\mu}\widehat{\gamma}_{\nu}
\widehat{\gamma}_{\alpha}\left[D_{\beta}^{\omega}+i\widehat{\gamma}_{\beta}\frac{\left(2+\zeta \right)\gamma_{5}}{4\,l}\right]\Psi. \hspace{0.6cm}
\end{align}
Now, due to the following identity
\begin{eqnarray}
\gamma_{a}\gamma_{b}\gamma_{c}=\eta_{ab}\gamma_{c}+\eta_{bc}\gamma_{a}- \eta_{ca}\gamma_{b}+i \,\epsilon_{abcd}\, \gamma_{5}\gamma^{d},
\end{eqnarray}
we can rewrite the totally anti-symmetric product between tetrads in (\ref{equations}) as follows
\begin{eqnarray}\label{product}
\epsilon^{\mu\nu\alpha\beta}\,\gamma_{5}
\widehat{\gamma}_{\mu}\widehat{\gamma}_{\nu}
\widehat{\gamma}_{\alpha}=\epsilon^{\mu\nu\alpha\beta}\gamma_{5}\gamma_{a}\gamma_{b}\gamma_{c}\,e_{\mu}^{a}\,e_{\nu}^{b}\,e_{\alpha}^{c}=\nonumber \\
i\,\gamma^{d}\,\epsilon^{\mu\nu\alpha\beta}\epsilon_{abcd}\,e_{\mu}^{a}\,e_{\nu}^{b}\,e_{\alpha}^{c},\hspace{1.5cm}
\end{eqnarray}
where we have used the relation between the (symmetric) metric tensor $g_{\mu\nu}$ and tetrads.
The inverse of a tetrad $e_{d}^{\beta}$ can be written as \cite{Westman}
\begin{eqnarray}\label{product}
e_{d}^{\beta}=-\frac{1}{3!\,|e|}\,\epsilon^{\mu\nu\alpha\beta}\epsilon_{abcd}\,e_{\mu}^{a}\,e_{\nu}^{b}\,e_{\alpha}^{c},
\end{eqnarray}
where $|e|=-\frac{1}{4!}\,\epsilon^{\mu\nu\alpha\beta}\epsilon_{abcd}\,e_{\mu}^{a}\,e_{\nu}^{b}\,e_{\alpha}^{c}\,e_{\beta}^{d}$ is the determinant of tetrads, having $\epsilon^{\mu\nu\alpha\beta}\epsilon_{\mu\nu\alpha\beta}=-4!$ as convention.
Thus, after a rescaling of the spinor field, $\Psi\rightarrow \psi=\Psi/\sqrt{l^{3}}$ and $\overline{\Psi}\rightarrow \overline{\psi}=\overline{\Psi}/\sqrt{l^{3}}$, we can easily see that the corresponding action becomes the following one
\begin{eqnarray}
S_{Dirac}\left[\overline{\psi},\psi, \omega_{\mu}\right]=\int_{M_{4}} d^{4}x\,|e|\,
\overline{\psi}\,(i\,\widehat{\gamma}^{\beta}D_{\beta}^{\omega}-\gamma_{5}m_{C})\psi,
\end{eqnarray}
which is nothing but the Dirac action with a dimensioful Dirac spinor $\psi$, where $m_{C}=(2+\zeta)/l$ is the chiral mass.
Note that the hermitian conjugate of the kinematic term in this action can be derived within our formalism in a straightforward way.
It is important to remark the fact that the chiral mass is invariant under chiral transformations but most of the known massive fermions in the Standard Model possess only a Dirac mass.
However, $m_{C}$ appears, for example, in some effective Dirac Hamiltonians that describe topological insulators and superconductors \cite{Ryu}.\\ Finally, note that the other terms in (\ref{spinor-BF}) proportional to $\omega_{\mu}$, $R_{\mu\nu}^{\omega}$ and $T_{\mu\nu}$ cannot generate any dynamical theory and disappear in the flat limit.

\section{Gravity, gauge fields and four-fermion interactions}
\label{Gravity, gauge fields and four-fermion interactions}
In this section, we consider our model in a more general context, where gravity, other gauge fields and suitable four-fermion interactions are taken into account.\\
Firstly, it is well known that in the first-order formalism, the gravitational action is written in terms of spin connection and tetrads. When they are considered independent variables, then a non-null torsion is allowed \cite{Hehl}. Moreover, dynamical fermionic matter, added to gravitational theory, becomes the natural source of torsion.
Consequently, it is quite natural to try to deal with gravity and fermions on the same footing. This is indeed possible because the Einstein-Cartan theory with a positive cosmological constant can be reformulated as a gauge-like theory with a broken de Sitter symmetry as discovered by MacDowell and Mansouri \cite{MM}. Its underlying geometric structure is naturally described within Cartan-geometry formalism \cite{Wise}. Importantly, an alternative derivation of the MacDowell-Mansouri action from a topological theory has been recently proposed in Ref. \cite{Palumbo-Gravity}. Thus, following our approach, gravity and fermions can be analyzed within a common geometric framework.\\
Furthermore, we can easily couple the spinor field in (\ref{spinor-BF}) with other gauge fields via the following replacements
\begin{eqnarray}\label{gauge}
D_{\beta}^A\rightarrow D_{\beta}^{A}+V_{\beta},
\end{eqnarray}
where $V_{\beta}$ is a generic vector field. It is straightforward to see that the corresponding twisted Dirac operator $\displaystyle{\not}D^{g}$ in the flat spacetime gets the standard form
\begin{eqnarray}
\displaystyle{\not}D^{g}=i\gamma^{\beta}\partial_{\beta}+i\gamma^{\beta}V_{\beta}.
\end{eqnarray}
At the same time, suitable four-fermion interactions \cite{NJL} can be included by adding to $S_{DFD}$ the following quartic term
\begin{eqnarray}\label{four-fermion}
S_{int}=\sigma\int_{M_{4}} d^{4}x\,
\,\epsilon^{\mu\nu\alpha\beta}  (\overline{\Psi}\,F_{\mu\nu}\Psi)(\overline{\Psi}\,F_{\alpha\beta}\Psi),
\end{eqnarray}
where $\sigma$ is a dimensionless parameter. This is the only possible quartic term that we can build through the Levi-Civita symbol and $F_{\mu\nu}$ without introducing any metric tensor. In the flat spacetime this term becomes
\begin{eqnarray}
\epsilon^{\mu\nu\alpha\beta}(\overline{\psi}\,[\gamma_{\mu},\gamma_{\nu}]\psi)(\overline{\psi}\,\,[\gamma_{\alpha},\gamma_{\beta}]\psi),
\end{eqnarray}
and deserves further studies, because is not included in the standard Nambu-Jona-Lasinio model \cite{NJL}.

\section{Conclusions}
\label{Conclusions}
To summarize, in this work we have introduced a de Sitter-gauge-invariant fermion theory in curved spacetime that has no local propagating degrees of freedom, i.e. it is topological. For this reason, we have dubbed it spinor topological field theory.
We have shown that the dynamics of spinor field emerges when the de Sitter gauge invariance breaks down to the Lorentz gauge one, giving rise to the Dirac theory in the curved spacetime. 
Consequently, within this formalism, we have also provided a geometric origin to the fermion mass. Thus, our mechanism combined to the standard Higgs mechanism in the electroweak interaction could give rise to the values of some fermion masses in the Standard Model.
Finally, we have also shown that other gauge fields and suitable four-fermion interactions can be easily included in our theory. We will analyze the details of the symmetry-breaking mechanism in future work.

\section{Acknowledgments}
\label{Acknowledgments}

We thank M. Cirio and an anonymous referee for insightful comments.




\end{document}